\documentclass[a4paper,10pt]{article}

\usepackage[utf8]{inputenc}
\usepackage[T1]{fontenc}
\usepackage{xcolor,graphicx}
\usepackage[top=0.6in,bottom=0.6in,right=1in,left=1in]{geometry}

\usepackage{natbib}
\usepackage[english]{babel}
\usepackage{hyperref}

\newcommand\aap{A\&A}                
\newcommand\aaps{A\&AS}              
\newcommand\apj{ApJ}                 
\newcommand\apjl{ApJ}                
\newcommand\mnras{MNRAS}             
\newcommand\pasa{Publ. Astron. Soc. Australia}  
\usepackage{authblk}
\usepackage{multicol}
\usepackage{graphicx}
\usepackage{amsmath}

\definecolor{blue-violet}{rgb}{0.54,0.17,0.89}
\begin{document}
\begin{titlepage}
\begin{center}
	\begin{minipage}{15cm}
	\begin{center}
		\includegraphics[width=2cm,height=2.2cm]{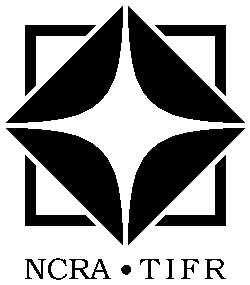}
	\end{center}
\end{minipage}\hfill\hfill\vspace{1cm}
	\begin{center}
	\textbf{\Large \color{blue-violet} National Centre for Radio Astrophysics}\\[0.1cm]
	\textbf{\large \color{blue-violet}  Tata Institute of Fundamental Research}
	\end{center}
\vspace{0.6cm}
\small Internal Technical Report: R1401

\vspace{5cm}


{ \Large \textbf{The Polarization Convention of the uGMRT in Band 4\\[0.4cm]} }

{\large Barnali Das, Sanjay Kudale, Poonam Chandra, Bhaswati Bhattacharya, Jayanta Roy, Yashwant Gupta}\\[0.5cm]
\textit{Corresponding Author's email id: \textsf{\color{blue}barnali@ncra.tifr.res.in} }\\[1cm]
 April 18, 2020\\

\vspace{5cm}
\begin{minipage}{20cm}
{\large \textbf{Objective: To check the polarization configuration of the uGMRT in band 4}} \\[0.5cm]
\end{minipage}{}

\vfill


\end{center}
\end{titlepage}
\normalsize

\section{Introduction}\label{sec:intro}
In astronomy, the polarization of a celestial signal is expressed in terms of the four Stokes parameters: $I$, $Q$, $U$ and $V$; where $I$ is the total intensity, $Q$ and $U$ represent linearly polarized power and $V$ quantifies the amount of circular polarization. These four quantities are derived from the voltages (coming from the celestial object) recorded by two receivers in the telescope corresponding to two orthogonal polarizations: $X$ and $Y$, for telescopes with dipolar feeds; or $R$  and $L$, for telescopes with circular feeds. Because of the degeneracy involved in forming a set of orthogonal polarizations, a convention is adopted by the IAU to define a set of universal $X$ and $Y$: in this convention, $X$ and $Y$ point towards the north and east respectively and the $+Z$ axis is the direction along which the radiation propagates from the source towards the observer \citep[e.g.][]{hamaker1996}. In this convention, $R$ and $L$ are related to $X$ and $Y$ as $R=(X+iY)/\sqrt{2}$, $L=(X-iY)\sqrt{2}$, $i=\sqrt{-1}$. The four Stokes parameters are then obtained using:
\begin{align}
    I&=XX^*+YY^*\equiv RR^*+LL^*\label{eq:I}\\
    Q&=XX^*-YY^*\equiv RL^*+R^*L\label{eq:Q}\\
    U&=XY^*+X^*Y\equiv -i(RL^*-R^*L)\label{eq:U}\\
    V&=-i(XY^*-X^*Y)\equiv RR^*-LL^*\label{eq:V}
\end{align}
The asterisk implies complex conjugate. Note that these four equations are w.r.t. the incoming radiation (and not what is received by the feed since the polarization state `seen' by the feed can be different from the intrinsic state if the feed is at the primary or tertiary focus of the antenna dish). Out of these four equations, there is a lot of confusion about the fourth equation that defines Stokes $V$. After defining $X$ and $Y$ according to the IAU convention and taking $R=(X+iY)/\sqrt{2}$, $L=(X-iY)\sqrt{2}$; $RR^*$ and $LL^*$ correspond to right and left hand circular polarization respectively. A right hand circularly polarized radiation is the one where for the incoming radiation, the electric field vector rotates counter-clockwise. Similarly, for left hand circularly polarized radiation, the electric field vector rotates clockwise for the incoming radiation \citep{robishaw2018}. This way of defining right and left hand circular polarization corresponds to the IEEE convention, also adopted by the IAU. Stokes $V$ is then positive for right hand circularly polarized radiation under the ``IAU/IEEE'' convention. However the alternate way of defining Stokes V, i.e. $V=LL^*-RR^*$ is also prevalent, especially among the pulsar community. This convention, where Stokes $V$ is positive for left hand circularly polarized radiation, is known as the ``PSR/IEEE'' convention.


In this report, we aim to find out the polarization convention adopted by the upgraded Giant Metrewave Radio Telescope \citep[uGMRT,][]{gupta2017} in band 4 (550--900 MHz). The feeds in this band are dipolar; however these are converted to circular polarizations and the final products that an observer can get are $RR^*$, $LL^*$ and the cross terms of $R$ and $L$. The Stokes parameters can then be obtained using Eq.s \ref{eq:I} to \ref{eq:V}.

This report is structured as follows: in \S\ref{sec:strategy}, we describe the strategy adopted to perform the experiment; this is followed by observation and data analysis (\S\ref{sec:data_analysis}). We present our results in \S\ref{sec:results}, and then compare those with the conventions adopted for the Very Large Array (VLA) telescope in \S\ref{sec:vla_comparison}. We present our conclusion in \S\ref{sec:conc}.
\section{Strategy}\label{sec:strategy}


The strategy that we adopted was to observe a source with known polarizations with the uGMRT and then compare the result with its known values. We choose the pulsar B1702--19 (RA: $17^h05^m36^s.099$ Dec: $-19^d06^m38^s.60$) as a test source to achieve this goal. 
This pulsar was chosen as it shows high circular polarization ($\approx 40\%$) according to the EPN database (\url{http://www.epta.eu.org/epndb/#gl98/J1705-1906/gl98_610.epn}) at 610 MHz. The Stokes $V$ sign is predominantly negative. This property is crucial if we want to detect circular polarization by imaging as it involves averaging over time. The convention used for Stokes $V$ was $V=LL^*-RR^*$ \citep[obtained by ][using the Lovell telescope]{gould1998}. Comparison of the 1369 MHz Stokes V profile for PSR B1702--19 obtained by \citet{johnston2017} using the Parkes radio telescope (explicitly mentioning the PSR/IEEE convention) with that obtained by \citet{gould1998}  using the Lovell telescope at 1408 MHz indicate that convention adopted by Lovell telescope is the PSR/IEEE version.
This implies that at 610 MHz, the pulsar is $\approx 40\%$ right circularly polarized according to the IEEE convention.

\begin{figure}
\centering
\includegraphics[width=0.45\textwidth]{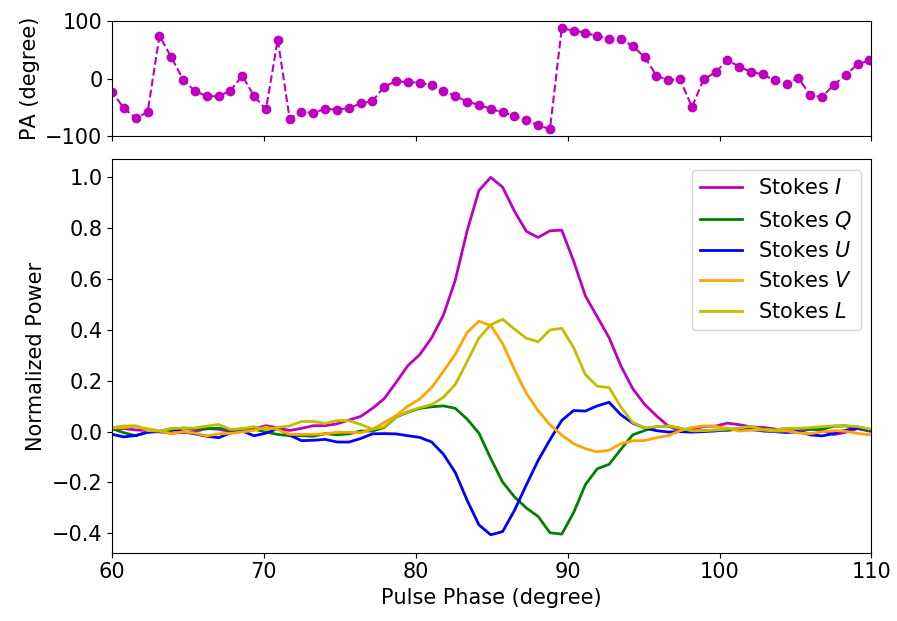}
\caption{The 610 MHz profiles for the pulsar B1702--19 for Stokes $I$ (magenta curve), $Q$ (green curve), $U$ (blue curve) and $V$ (orange curve) obtained by \citet{gould1998} using the Lovell telescope. Also shown are the angle of polarization (PA) on top panel and linear polarization $L=\sqrt{Q^2+U^2}$ in yellow in the bottom panel. Note that the convention for Stokes $V$, here, is $V=RR^*-LL^*$ even though the convention used in the original paper \citep{gould1998} was $V=LL^*-RR^*$. This is done to avoid confusion since GMRT convention for defining $V$ is $RR^*-LL^*$. For details, refer to \S\ref{sec:strategy}.\label{fig:epn_profile}}
\end{figure}

In Figure \ref{fig:epn_profile}, we show the 610 MHz $I$, $Q$, $U$, $V$ profiles for this pulsar, generated from the data available in the EPN database \citep{gould1998}. To make it consistent with the rest of the paper, we have plotted $V=RR^*-LL^*$, in stead of the original convention of $V=LL^*-RR^*$.



\section{Observation and data analysis}\label{sec:data_analysis}
We observed this pulsar in band 4 of the uGMRT on March 19, 2020 with 200 MHz bandwidth, divided into 2048 channles. 3C286 was used as the flux and bandpass calibrator, and J1822-096 (RA: $18^h22^m28.71^s$, Dec: $-09^d38^m56^s.84$) was used as the phase calibrator. The on-source time was around 40 minutes. The data were recorded both in interferometric and pulsar phased array mode. 

In the pulsar mode, the array was first phased using the nearby calibrator J1822-096. The data were then recorded on target in GMRT format (co- and cross-polar voltage products, viz. $RR^*$, $LL^*$, and real and imaginary parts of $RL^*$) with 
a time resolution of 327.68 microseconds ($\approx 0.33$ ms, the rotation period of the pulsar is $\approx 280$ ms).
These data were analysed using the full polarization data analysis pipeline of the GMRT described in \citet{kudale2008}. $R$ and $L$ powers are separately scaled by its average bandshape to take out the bandpass 
effect. The final outputs were Stokes $I$, $Q$, $U$ and $V=RR^*-LL^*$, which are normalized by the maximum in total power (i.e. Stokes-I). 
Note that no polarization calibration was applied.

The interferometric data were flagged and calibrated using standard tasks in $\textsc{casa}$ \citep[e.g.][]{das2019}. The $RR^*$ and $LL^*$ data of the target were separately self-calibrated to make the final images.

\section{Results and discussions}\label{sec:results}
\begin{figure*}
\centering
\includegraphics[width=0.4\textwidth]{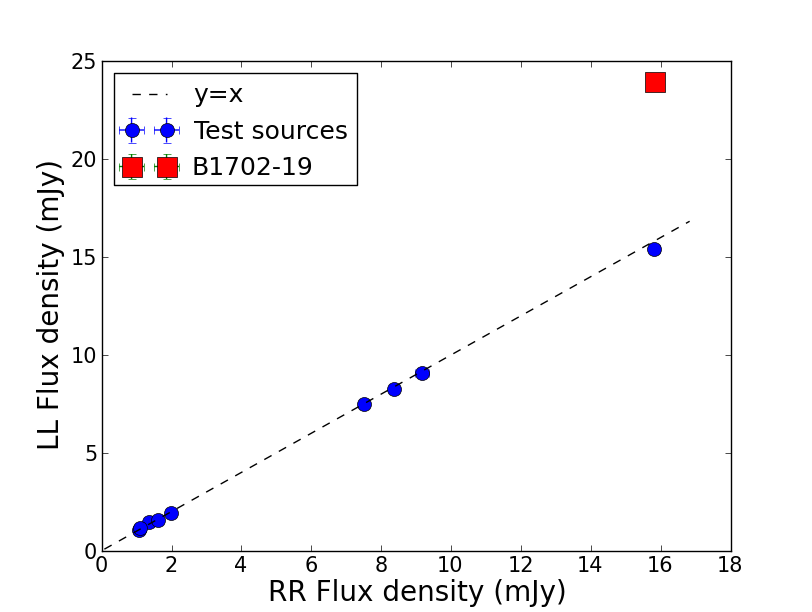}
\includegraphics[trim={3.2cm 2cm 3.8cm 3.1cm}, clip, width=0.5\textwidth]{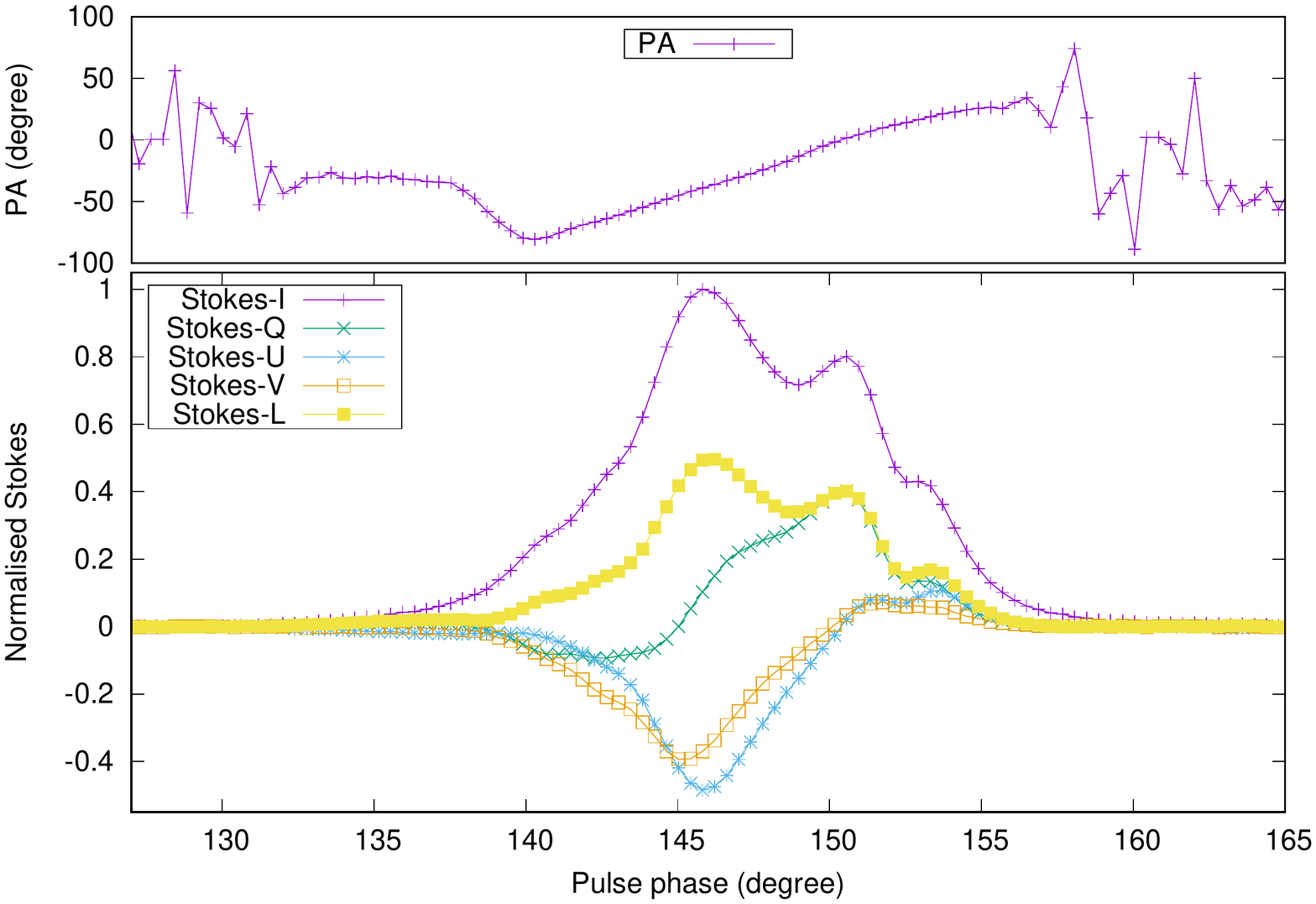}
\caption{\textit{Left:} Comparion of $RR^*$ and $LL^*$ flux densities for the pulsar B1702--19, obtained by imaging, and that for a few test sources in the field of view in band 4 (550--750 MHz) of the uGMRT. \textit{Right:} The Stokes $I$ (magenta curve), $Q$ (green curve), $U$ (blue curve) and $V=RR^*-LL^*$ (orange curve) profiles for the pulsar B1702--19 at 650 MHz as measured by the uGMRT. Also shown is the linear polarization $L=\sqrt{Q^2+U^2}$ in yellow.  \label{fig:band4_ll_rr}}
\end{figure*}

We present the results obtained from the data taken in  interferometric mode and pulsar (beamformer) mode in the following two subsections.

\subsection{Results from interferometric mode}\label{subsec:interferometry_mode}
In the left panel of Figure \ref{fig:band4_ll_rr}, we show the result of the imaging exercise, where we plot the flux denities in $LL^*$ against the flux densities in $RR^*$ for the target (marked in red square) as well as a few other sources (marked in blue circles) in the field of view. This was done to make sure that the inferred difference in flux densities is not due to any systematics. We clearly see that for these test sources, the flux densities nearly follow the $y=x$ line, meaning that they have equal flux densities in $RR^*$ and $LL^*$. But the target is significantly offset from the $y=x$ line. Its $RR^*$ flux density came out to be $15.8\pm 0.1$ mJy, whereas the $LL^*$ flux density came out to be $23.9\pm 0.2$ mJy which give percentage circular polarization $V/I=-20\%$, while using $V=(RR^*-LL^*)/2$ and $I=(RR^*+LL^*)/2$.

Thus, according to the uGMRT data, the pulsar is left hand circularly polarized at 610 MHz, whereas it is right hand circularly polarized under the IEEE convention. Therefore, effectively the uGMRT $RR^*$ is actually $LL^*$ in IEEE convention and vice-versa in band 4.

\subsection{Result from pulsar mode}\label{subsec:pulsar_mode}
In the right panel of Figure \ref{fig:band4_ll_rr}, we show the pulse profiles for Stokes $I$, $Q$, $U$ and $V\,(=RR^*-LL^*)$. The maximum circular polarization observed is $-40\%$. By comparing these profiles with those in Figure \ref{fig:epn_profile}, we find that:
\begin{enumerate}
    \item The signs of the Stokes $Q$ profiles in the two figures do not match, but the Stokes $U$ profiles agree with each other.
    \item The sign of the uGMRT Stokes $V$ profile does not match with that of the EPN Stokes $V$ profile.
    \item The sweeps of the polarization angle (PA) are opposite in the two figures.
\end{enumerate}
The third point is a consequence of the first point. The second point suggests that the uGMRT $RR^*$ and $LL^*$ in band 4 are in opposite conventions to that defined by the IEEE (also suggested by the data taken in interferometric mode). However simply swapping $R$ and $L$ cannot explain why the sign of $Q$ is opposite to what should have been obtained according to the IEEE convention, whereas signs of $U$ match. To explain this behaviour, we refer to Eq. \ref{eq:Q} and Eq.\ref{eq:U}: 
\begin{align}
    Q=RL^*+R^*L,\quad U=-i(RL^*-R^*L)\nonumber
\end{align}
Thus, swapping of $R$ and $L$ will give us opposite sign of $U$ and not $Q$. To explain the reversal of the sign of $Q$, we next consider the possibility that $X$ and $Y$ are swapped (\S\ref{sec:intro}). For clarity, let us denote uGMRT $R$ and $L$ as $\tilde{R}$ and $\tilde{L}$ respectively. If $X$ and $Y$ are swapped, we will get:
\begin{align*}
    \tilde{R}&=\frac{Y+iX}{\sqrt{2}}=i\frac{X-iY}{\sqrt{2}}=iL\\
    \tilde{L}&=\frac{Y-iX}{\sqrt{2}}=-i\frac{X+iY}{\sqrt{2}}=-iR
\end{align*}
In that case, we will get $\tilde{V}=\tilde{R}\tilde{R^*}-\tilde{L}\tilde{L^*}=LL^*-RR^*$, so that the Stokes $V$ profile obtained with the uGMRT will match the one from EPN even though they have different conventions. For $Q$ and $U$, we will get:

\begin{align*}
    \tilde{Q}&=\tilde{R}\tilde{L}^*+\tilde{R}^*\tilde{L}=(iL)(iR^*)+(-iL^*)(-iR)=-(RL^*+R^*L)=-Q\\
    \tilde{U}&=-i(\tilde{R}\tilde{L}^*-\tilde{R}^*\tilde{L})=-i\{(iL)(iR^*)-(-iL^*)(-iR)\}=-i(RL^*-R^*L)=U
\end{align*}

Thus, we infer that indeed the $X$ and $Y$ dipoles are swapped so that the resulting $RR^*$ and $LL^*$ are not in accordance with the IEEE convention in band 4 of the uGMRT.

\section{Comparison with the Very Large Array (VLA) telescope}\label{sec:vla_comparison}
To compare the definitions of circular polarization for the uGMRT and the VLA, we again used the same pulsar. According to the EPN database, the pulsar is predominantly right hand circularly polarized with $\approx 32\%$ circular polarization at 1.4 GHz \citep{gould1998}.

We analyzed archival VLA data for this pulsar acquired on February 9, 2002 at 1.5 GHz (project ID: AC629). The on-source time was around 11 minutes. No flux calibrator was observed. The phase calibrator was J1733-130 (J2000 RA: $17^h33^m02.70^s$, Dec: $-13^d04^m49^s.55$). Despite the lack of a flux calibrator, we did the analysis assuming a flat spectrum for the phase calibrator (with flux density of 1 Jy) over the observing band which was of width 22 MHz (fractional bandwidth is $1.5\%$). This implies that the flux densities that we get for the target could be incorrect, however the sign of Stokes $V$ will not be affected, and so is the case for the percentage circular polarization ($V/I$). The percentage circular polarization came out to be $+12\%$, while using $V=(RR^*-LL^*)/2$. 

We analyzed one more VLA archival dataset for this pulsar, obtained on February 5, 1990 (project ID: AR218) at 1.4 GHz with 100 MHz bandwidth (fractional bandwidth $\approx 7\%$). The on-source time was around 2 hours and 37 minutes. 3C286 was used as the flux and bandpass calibrator, and J1743-038 (J2000 RA: $17^h43^m58^s.86$, Dec: $-03^d50^m04^s.62$) was used as the phase calibrator. We obtained the flux densities of the target to be $7.7\pm 0.4$ mJy in $LL$ and $10.0\pm 0.4$ mJy in $RR$, which again gives  $V/I\approx +13\%$.

Thus, according to the VLA convention also, the pulsar is right circularly polarized, implying that the $RR^*$ and $LL^*$ for the VLA are defined in accordance with the IEEE convention.

Note that the above result need not apply to the upgraded VLA, which is the Karl G. Jansky VLA or the JVLA. However we have indirect evidence in support of the assumption that the old VLA and the JVLA obey the same convention for defining $RR^*$ and $LL^*$. This comes from the observation of the magnetic star CU\,Vir. This is the first main-sequence star which was discovered to be a persistent emitter of electron cyclotron maser emission (ECME) at 1.4 GHz \citep{trigilio2000}. This result came out from an observation performed with the VLA in the year 1998. One characteristic of ECME is that it produces very highly circularly polarized pulsed emission. For CU\,Vir, the ECME pulses were found to be right circularly polarized \citep{trigilio2000}. The star was again observed with the VLA in the year 2010 at 1.45 GHz, and the ECME pulses were found to be right circularly polarized \citep{trigilio2011}. Most recently, we observed the star with the JVLA in the year 2019, and we also observed that the pulses are right circularly polarized in both L (1--2 GHz) and S (2--4 GHz) bands (Das et al. in prep). This suggests that the JVLA convention for defining $RR^*$ and $LL^*$ is same as that for the old VLA, and hence in accordance with the IEEE convention.

\section{Conclusion}\label{sec:conc}
The primary conclusion drawn from this work is the following:
\begin{enumerate}
\item The $X$ and $Y$ dipoles of the uGMRT in band 4 are swapped (w.r.t. the IAU convention).

\item If we have to compare uGMRT band 4 polarization results with that from the VLA, Lovell or Parkes telescope, we must interchange $RR^*$ and $LL^*$ for the uGMRT data; and also change the sign of Stokes $Q$.
\end{enumerate}
We would like to mention that the observatory is likely to take measures to make the convention (for defining $X$ and $Y$) consistent with the standard convention. These new measures, once implemented and fully tested, will be notified in a future technical report. Until then, the above conclusions are valid.



\section*{Acknowledgements}
We thank the staff of the
GMRT that made these observations possible. The GMRT is run by
the National Centre for Radio Astrophysics of the Tata Institute of
Fundamental Research. BD thanks Minhajur Rahaman for suggesting the pulsar B1702--19 to use as the polarization standard. BD thanks Biny Sebastian and Jayaram N. Chengalur for useful discussions. SK thanks Kadaladi Pavankumar for helping in beam polarization data. PC thanks Dipanjan Mitra for independently suggesting the polarization standard. This research has used NASA's Astrophysics Data system.

\bibliographystyle{mnras}

\begin{thebibliography}{99}
\bibitem[Das et al.(2019)]{das2019} Das, B., Chandra, P., Shultz, M.~E., et al.\ 2019, \apj, 877, 123

\bibitem[Gould \& Lyne(1998)]{gould1998} Gould, D.~M., \& Lyne, A.~G.\ 1998, \mnras, 301, 235

\bibitem[Gupta et al.(2017)]{gupta2017} Gupta, Y., Ajithkumar, B., Kale, H.~S., et al.\ 2017, Current Science, 113, 707

\bibitem[Hamaker \& Bregman(1996)]{hamaker1996} Hamaker, J.~P., \& Bregman, J.~D.\ 1996, \aaps, 117, 161

\bibitem[Johnston \& Kerr(2018)]{johnston2017} Johnston, S., \& Kerr, M.\ 2018, \mnras, 474, 4629

\bibitem[Kudale (2008)]{kudale2008} Kudale, S. \ 2008, M.Sc. Thesis, Technical Report NCRA-T1400

\bibitem[Robishaw \& Heiles(2018)]{robishaw2018} Robishaw, T., \& Heiles, C.\ 2018, arXiv e-prints, arXiv:1806.07391

\bibitem[Trigilio et al.(2000)]{trigilio2000} Trigilio, C., Leto, P., Leone, F., Umana, G., \& Buemi, C.\ 2000, \aap, 362, 281

\bibitem[Trigilio et al.(2011)]{trigilio2011} Trigilio, C., Leto, P., Umana, G., et al.\ 2011, \apjl, 739, L10

\bibitem[van Straten et al.(2010)]{van_straten2010} van Straten, W., Manchester, R.~N., Johnston, S., et al.\ 2010, \pasa, 27, 104

\end{thebibliography}

\end{document}